\documentclass{article}
\usepackage{graphicx} 
\usepackage{subfigure}
\usepackage{float}
\usepackage[letterpaper, margin=1in]{geometry}
\usepackage{authblk}
\usepackage{caption}
\usepackage{tgheros} 
 
\usepackage[T1]{fontenc}
\usepackage{abstract}

\usepackage[superscript]{cite}
\makeatletter\renewcommand\@citess[1]{\textsuperscript{[#1]}}\makeatother  
\usepackage{tabularx}
\usepackage{hyperref}
\usepackage{cite}
\usepackage{booktabs}
\usepackage{multicol}

\title{Automatic treatment planning for radiotherapy: a cross-modality and protocol study}
\author[1,3]{Gregory Szalkowski}
\author[2]{Xuanang Xu}
\author[1]{Shiva Das}
\author[2]{Pew-Thian Yap}
\author[1,*]{Jun Lian}

\affil[1]{Department of Radiation Oncology, University of North Carolina, Chapel Hill, NC}
\affil[2]{Department of Radiology and Biomedical Research Imaging Center, University of North Carolina, Chapel Hill, NC}
\affil[3]{Department of Radiation Oncology, Stanford University, Stanford, CA}
\affil[*]{Corresponding author: jun\_lian@med.unc.edu}

\begin{document}
\maketitle

\begin{abstract}
\normalsize
\textbf{Purpose:} This study investigates the applicability of 3D dose predictions from a model trained on one modality to a cross-modality automated planning workflow. Additionally, we explore the impact of integrating a multi-criteria optimizer (MCO) on adapting predictions to different clinical preferences.

\textbf{Methods:} Using a previously created three-stage UNet in-house model trained on the 2020 AAPM OpenKBP challenge dataset (340 head and neck plans, all planned using 9-field static IMRT), we retrospectively generated dose predictions for 20 patients. These dose predictions were in turn used to generate deliverable IMRT, VMAT, and Tomotherapy plans using the fallback plan functionality in Raystation. The deliverable plans were evaluated against the dose predictions based on primary clinical goals. A new set of plans was also generated using MCO-based optimization with predicted dose values as constraints. Delivery QA was performed on a subset of the plans to assure clinical deliverability.

\textbf{Results:} The mimicking approach accurately replicated the predicted dose distributions across different modalities, with slight deviations in spinal cord and external contour maximum doses. MCO customization significantly reduced doses to OARs prioritized by our institution while maintaining target coverage. All tested plans met clinical deliverability standards, evidenced by a gamma analysis passing rate above 98\%.

\textbf{Conclusions:} Our findings show that a model trained only on IMRT plans can effectively contribute to planning across various modalities. Additionally, integrating predictions as constraints in an MCO-based workflow, rather than direct dose mimicking, enables a flexible, warm-start approach for treatment planning. Together, these approaches have the potential to significantly decrease plan turnaround time and quality variance, both at high resource medical centers that can train in-house models, and smaller centers that can adapt a model from another institution with minimal effort.
\end{abstract}

\begin{multicols}{2}
\section*{Introduction}
\begin{figure}[H]
    \centering
    \includegraphics[width=0.9\linewidth]{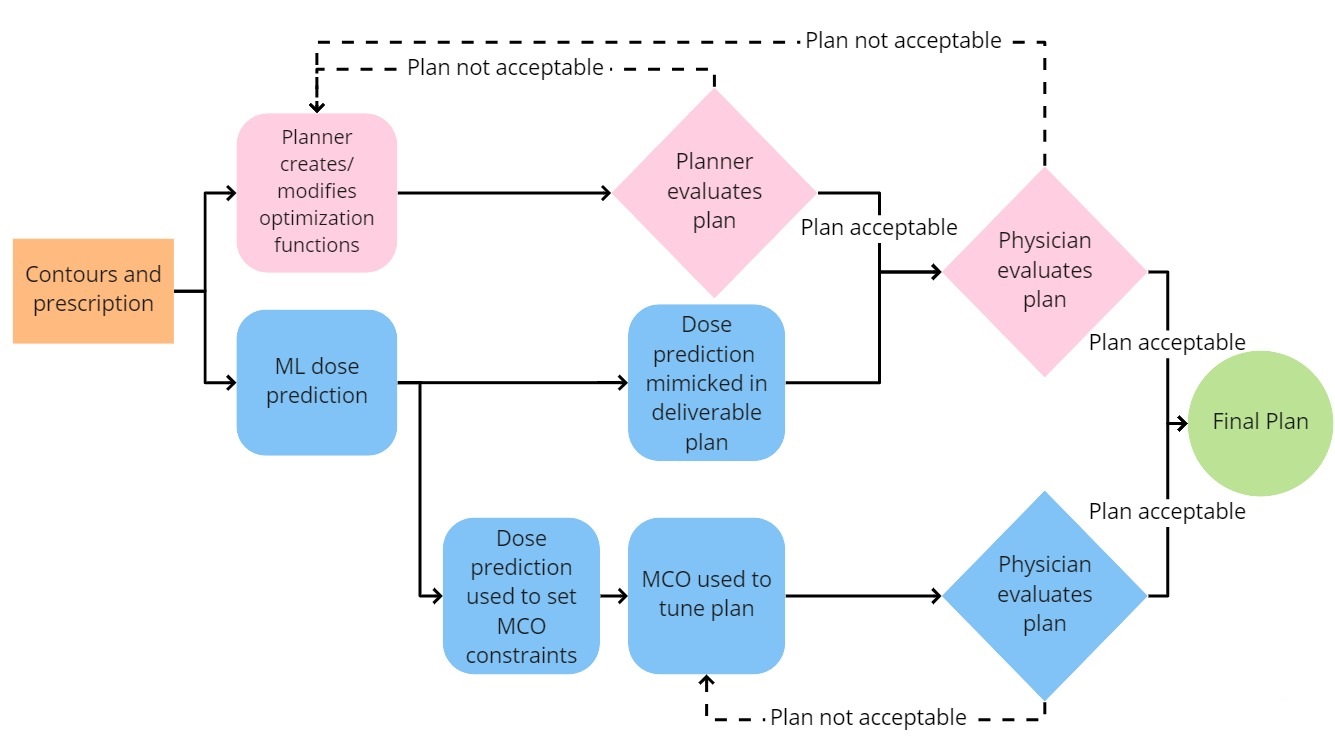}
    \caption{Conventional and proposed workflows. The conventional workflow (top, pink) requires a great deal of manual work, with the dosimetrist manually iterating through many possible plans. Our proposed workflows (bottom, blue) significantly cut down on this work by providing a prediction of what the dose distribution for a given patient would look like based on prior plans. While the mimic workflow may require some further optimization by the planner, the expected number of iterations is much less than in the conventional workflow.}
    \label{fig:workflow}
\end{figure}

\addtocounter{section}{1}
In the treatment of cancer, radiation therapy serves as a valuable tool and is estimated to provide a benefit for about half of all cancer patients\cite{Ref1}. For head and neck cancers, radiotherapy is indicated for an even higher percentage (approximately 74\%) of the patient population \cite{Ref2}. While many clinical trials have established suggested prescription doses as well as healthy tissue dose constraints for the treatment of these cancers, \cite{Ref3, Ref4,  Ref5} it may not always be possible to meet all these objectives for a given patient. In these cases, the planner and the physician must determine what trade-offs should be made to provide the best outcome for the patient. 
In standard planning the planner will set a number of optimization parameters, consisting of some dose objective and an associated weight, that the treatment planning system (TPS) then uses to create an optimized plan. Due to variations in patient anatomy and treatment intent the “ideal” optimization parameters are different for each case and the planner must iteratively adjust the parameters to try to improve the plan. This is complicated by the fact that the relationship between adjustments to the optimization parameters and the resulting changes to the optimized treatment plan are not always intuitive \cite{Ref6, Ref7}, resulting in a labor-intensive process, especially for planners with less experience. This means that the quality of the final treatment plans is often dependent upon the planner experience and available time for the planning process \cite{Ref8, Ref9, Ref10}, and low-quality plans can lead to worse clinical outcomes for patients \cite{Ref11, Ref12}. 
The use of automation can help with both reducing this variability in plan quality, as well as help increase patient throughput by speeding up the typically slow trial-and-error planning process \cite{Ref13, Ref14}. A variety of auto-planning methods have been developed \cite{Ref15, Ref16}, which can be roughly split into 1)  knowledge based planning, such as dose volume histogram (DVH) guidance, where DVHs for contoured structures are predicted based off of anatomical and geometric features \cite{Ref17, Ref18}, 2) protocol based optimization, where changes to the optimization parameters are automatically implemented to minimize organ at risk dose while meeting user defied clinical constraints\cite{Ref19, Ref20}, 3) automated multicriteria optimization, where the software generates a set of parieto-optimal plans and allows for either the user or the software itself to select the “best” solution based on the treatment site and clinical protocol \cite{Ref21, Ref22, Ref23}, and, most recently, 4) statistical models, including machine learning, which attempt to learn the correlation between patient anatomy and the resulting plan \cite{Ref24, Ref25}. Recent approaches have utilized deep learning for dose prediction \cite{Ref26, Ref27, Ref28, Ref29}, which could then be used to generate a plan via dose mimicking approximation or by directly predicting the fluence map that would produce the desired dose distribution \cite{Ref30, Ref31}.
In this work, we propose two automatic treatment planning methods, shown in Figure \ref{fig:workflow}, a mimicking approach, and a multi-criteria optimizer (MCO) approach. For the mimicking approach, we evaluate the applicability of the predicted dose distribution generated by our deep learning model to the achievable dose distribution using the same delivery modality as the model training set, fixed-gantry IMRT (FG-IMRT), as well as different delivery modalities, VMAT and Tomotherapy. Since the model will output a predicted dose distribution similar to the plans used for training, this approach is less suitable if a large variety of protocols, the directives stating the prescription dose to the target(s) and dose constraints of healthy tissue structures, are to be used. To address this, we investigate if integrating the multi-criteria optimizer into our workflow allows for modifications to be made to the plan dose distribution to better conform to institutional or individual physician preference without requiring the creation of a new model.
\end{multicols}

\begin{multicols}{2}

\section{Methods}
\subsection{Model Creation}
\begin{figure}[H]
    \centering
    \includegraphics[width=0.9\linewidth]{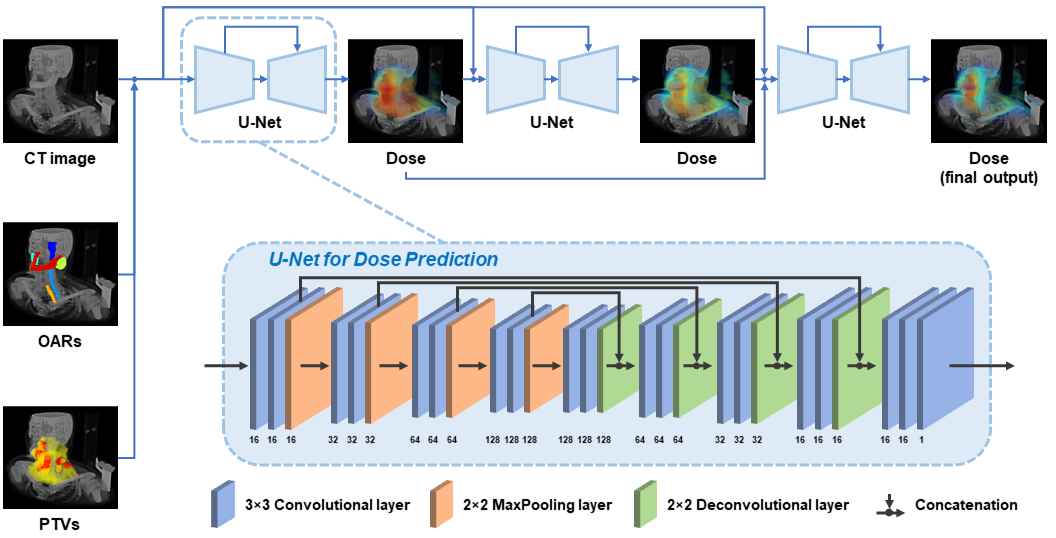}
    \caption{Schematic diagram of the network architecture of our dose prediction model
The bottom sub-figure illustrates the detailed structure of each 3D U-Net used for dose prediction. The numbers annotated below the layers indicate their output channel number.
}
    \label{fig:model}
\end{figure}

Our dose prediction model utilizes a triple-stage cascaded U-Net to predict the dose distribution based on the input CT image and the PTV/OAR structure set. The schematic diagram of our model is illustrated in Figure \ref{fig:model}. This model consists of three cascaded U-Nets, which sequentially predict the dose volume in a coarse-to-fine manner using the auto-context mechanism \cite{Ref32}. Specifically, the first U-Net takes the CT image and the ROI contours (OAR \& PTV) as input and outputs a coarse dose volume. This coarse dose volume is then fed into the second U-Net together with the CT image and the ROI contours to predict a dose volume that is more accurate than the first one. Finally, the third U-Net takes the two dose volumes predicted by the first and second U-Net together with the CT image and the ROI contours to generate the final dose volume, which is expected to be a refinement of the previous two dose volumes. The three U-Nets have identical network structures (which is illustrated in the bottom of Figure 2 in detail) except the input channel number of the first convolutional layer, which varies across the number of the input dose volumes. The network is fully implemented in 3D (which means it takes the 3D CT volume as input, not 2D CT slices). To assist with the training, we normalized the CT values from [-600, 1400] HU to [0, 1] and the dose values from [0, 80] Gy to [0, 1]. We trained the model using Adam optimizer with a base learning rate of 0.0005 for 400 epochs (approximately 17 hours) with a batch size of 6. The training procedure was conducted on a server computer equipped with two Intel(R) Xeon(R) E5-2650 CPUs working at 2.20GHz and six NVIDIA TITAN Xp graphic cards with 12 GBytes of memory each. We implemented our model using the PyTorch framework.
\end{multicols}

\begin{multicols}{2}

We used the dataset provided for the 2020 AAPM (American Association of Physicists in Medicine) OpenKBP challenge \cite{Ref33, Ref34}, consisting of 340 head and neck plans which all utilized 9 field IMRT. Each plan had either just a high risk PTV, a high risk and a low risk PTV, or a high, intermediate, and a low risk PTV. The dose prescription for the PTVs was 70 Gy, 63, and 56 Gy for the high, intermediate, and low risk volumes, respectively. 200 plans were used for training, 40 for validation, and 100 for testing. The model performance was one of the best among 195 registered participants, according to the result on the official challenge closing date May 29th, 2020.

\subsection{Same Modality Assessment}
After generating the dose predictions using our machine learning model, we imported the predicted dose, CT, and RT structure data for 20 patients into our treatment planning system, Raystation (RaySearch Laboratories, Stockholm, Sweden). We then set up a plan template with the same beam arrangement as the plans used in the training set; 9 IMRT beams from 0 to 320 degrees, with 40 degrees between each beam. The beam isocenter was set using the BBs placed on the patient’s skin, if available, or at the center of the high risk PTV. We then set the beam optimization parameters to our institutional standard to assure the resulting plans would be deliverable on our machines. For IMRT plans, the minimum segment area was restricted to 4 cm2, the minimum segment MU per fraction was set to 5 MU, and the minimum leaf end separation was set to 1 cm. 
Using Raystation’s mimic dose functionality, we optimized the plan for each patient to try to match the dose prediction. For each case, the optimization was run for 180 iterations or until convergence, whichever happened first. All the plans were normalized to give 95\% coverage at 70 Gy to the high risk PTV, per our institutional standard. The resulting plan dose was then compared against the predicted dose using the methods laid out in section 2.6.

\subsection{Cross Modality, fixed gantry IMRT (FG-IMRT) to VMAT}
To assess if the predicted dose could be extended to treatment modalities other than those used in the plans in the training set, we also applied the mimic dose technique to create two arc VMAT plans, which is the primary modality used at our clinic. For the VMAT plans, the leaf motion was constrained to 0.48 cm/degree.
While the plan template was different in these cases, the rest of the process was the same as the same modality plans; a deliverable plan was optimized using the mimic dose functionality, using the predicted dose distribution as the reference.

\subsection{Cross Treatment machine, FG-IMRT to Tomotherapy}
Finally, we also assessed if the predicted dose could be extended to a different delivery device entirely, a Tomotherapy unit. For the Tomotherapy plans, we used a dynamic field width of 2.5 cm, a pitch factor of 0.303, and a delivery time factor max of 1.5, the standard settings used for head and neck planning at our institution.

\subsection{Dosimetric protocol}
To allow for more customization of the plan to conform to differences in preference between institutions and individual physicians and to help improve plan quality when the prediction is not optimal, we then used the dose prediction to inform the use of the MCO to create deliverable plans. In brief, for a given set of optimization objectives, the MCO generates a database of Pareto optimal plans which sample the Pareto surface consisting of all possible Pareto optimal plans\cite{Ref31, Ref35, Ref36, Ref37}. By navigating this Pareto surface, it is possible to explore a wide range of tradeoffs to search for the “best” plan, though in practice a large area of this surface concerns tradeoff that would be considered clinically unacceptable. By applying constraints set using the knowledge gained from our dose prediction, we can focus the MCO on the area of the Pareto surface containing what would be considered high quality, reducing the necessary computational time and increasing the ease of navigation. 
In this work, for each case, a standard template of MCO objectives and constraints (Supplementary Table \ref{tab:sup_tab_1}) was loaded, and the constraints were updated to correspond to the results from the dose prediction for the patient. During the MCO optimization, Raystation uses the objectives to create the set of pareto plans and will only consider pareto plans that meet all specified constraints. Since we wanted to leave some flexibility for customization and prevent the MCO from becoming over constrained, we elected to only set some of the predicted clinical goals as constraints. As will be discussed in Sections 3.1-3.3, the mimic dose functionality had some difficulty replicating the maximum dose to the spinal cord and the external contour, so these were chosen to be set as constraints in the MCO template. We also set the D50\%  for the parotids as a constraint, as due to their proximity to the target the parotids were found to have the largest impact on the dose distribution. D50\% was used instead of mean as it was found to be less likely to lead to the MCO optimization becoming over constrained. Not all contours were present for all patients, namely intermediate risk PTV, esophagus, and larynx contours, so the associated objective functions were only used when applicable.
After the template was updated with the cases specific constraints from the dose prediction, we ran the MCO optimization using the Raystation recommended parameters; generating a number of pareto plans equal to twice the number of objectives used (24-30 in these cases, depending on the structures contoured) with a maximum of 40 iterations per pareto plan. In 2 cases, the max dose to external constraint violated the MCO feasibility constraint and was excluded. After the pareto plans were generated, the “balance plan,” which equally weights all the pareto plan results, was used without modification to generate the final deliverable plan. In clinical practice, it would be prudent to adjust the weighting of the pareto plans to improve the dosimetry of more concerned structures and/or conform to institutional preference, but this was not done in this work to maintain a standardized workflow. 
\subsection{Comparison Metrics}
The similarity between the predicted dose and the mimicked or MCO integrated plans was primarily assessed using the clinical goals established at our institution for the structures that were contoured in the OpenKBP challenge. These metrics consist of the max dose to the spinal cord and external, the mean dose to the parotids and larynx, and the D50 to the parotids. Supplementary Table \ref{tab:sup_tab_2} shows the template goal sheet used across the test cases. The statistical significance of the difference was assessed using a Wilcoxon signed rank test. We defined a dose difference > 5\% as “clinically relevant.” 

\subsection{Delivery Assessment}
To assure that the mimicked plans were deliverable on the machines, we performed delivery QA (DQA) on a selected sample of five plans for each modality (IMRT, VMAT, and Tomotherapy) for both the mimicked and MCO plans. To do this, we delivered the plans onto an ArcCheck device software (Sun Nuclear Corporation, Melbourne, FL) that measured the dose distribution output by the treatment machine. We then compared this measured distribution to the predicted distribution provided by our treatment planning system using a gamma analysis test\cite{Ref38}.
 We used 3\% for the dose-difference and 2mm distance-to-agreement criteria, a low-dose cut-off of 10\% of the prescription dose and considered a plan to be clinically deliverable if $\geq$ 90\% of points passed, as recommended by American Association of Physicists in Medicine task group report 218 \cite{Ref39}. The analysis was run using the SNC Patient software provided with the ArcCheck device. (Sun Nuclear Corporation, Melbourne, FL)

\section{Results}
\subsection{Same Modality Mimic Plans}
Except for the spinal cord maximum dose, there were no clinically relevant (>5\%) difference between the predicted and mimicked IMRT plans for the relevant dosimetric goals. An extra max dose objective on the cord was able to reduce the difference in the cord max dose below 2\% without a significant impact on any of the rest of the other clinical goals. As will be covered in the discussion, the addition of a simple maximum dose objective can assist the mimic workflow in achieving the predicted value for this clinical goal. Figure \ref{fig:predM_a} shows the distribution of the percent difference between the predicted doses and mimicked IMRT plans for select dosimetric values. A full comparison can be found in Supplementary Table \ref{tab:sub_tab_3}.

\subsection{Cross Modality from FG-MRT to VMAT}
For the mimic plans created using VMAT, the dosimetry for most organs is still similar to the prediction based on FG-IMRT model. There were no clinically relevant (>5\%) differences for any of the relevant clinical goals, though the difference in the whole body (also known as the external contour) and cord max dose was statistically significantly different (p<0.04) from the prediction. \ref{fig:predM_b} shows the distribution of the percent difference between the predicted doses and mimicked VMAT plans for select dosimetric values. A full comparison can be found in Supplementary Table \ref{tab:sup_table_4}.

\subsection{Cross Modality from FG-IMRT to Tomotherapy}
\begin{figure}[H]
    \centering
    \subfigure[Shows the difference between the prediction and the mimic 9 field IMRT plan. The central mark indicates the median, and the bottom and top edges of the box indicate the 25th and 75th percentiles, respectively. Any outliers, points more than 2.7 standard deviations away from the median, are plotted outside of the whiskers.]{
        \includegraphics[width=0.45\textwidth]{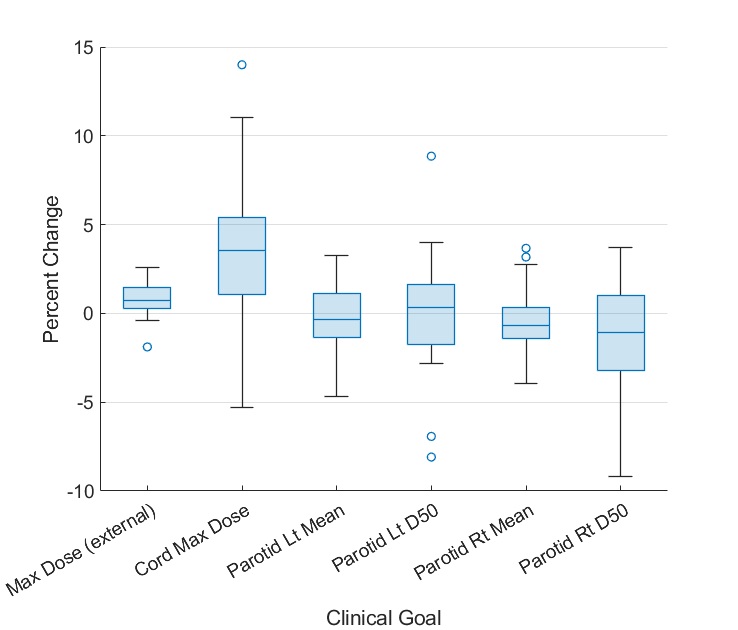}
        \label{fig:predM_a}
    }
    \subfigure[As a, but comparing the dual arc VMAT plans]{
        \includegraphics[width=0.45\textwidth]{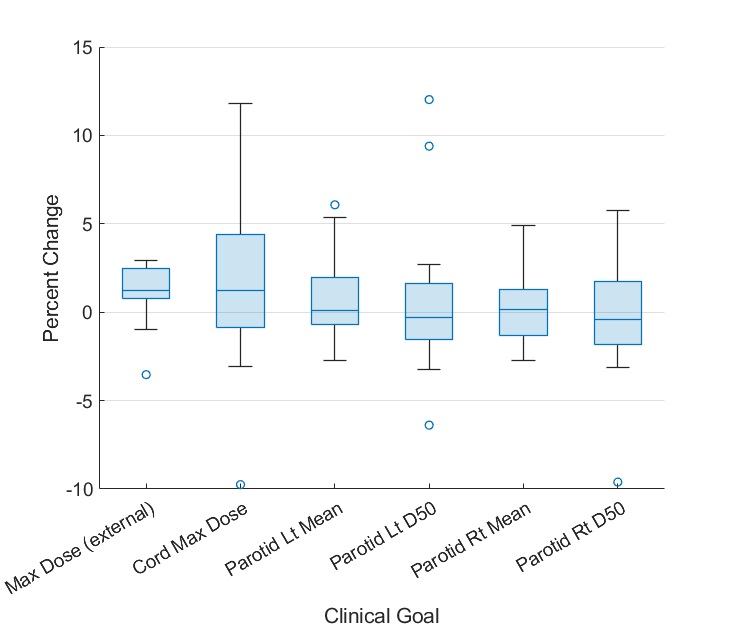}
        \label{fig:predM_b}
    }
    \subfigure[As a, but comparing the Tomotherapy plans]{
        \includegraphics[width=0.45\textwidth]{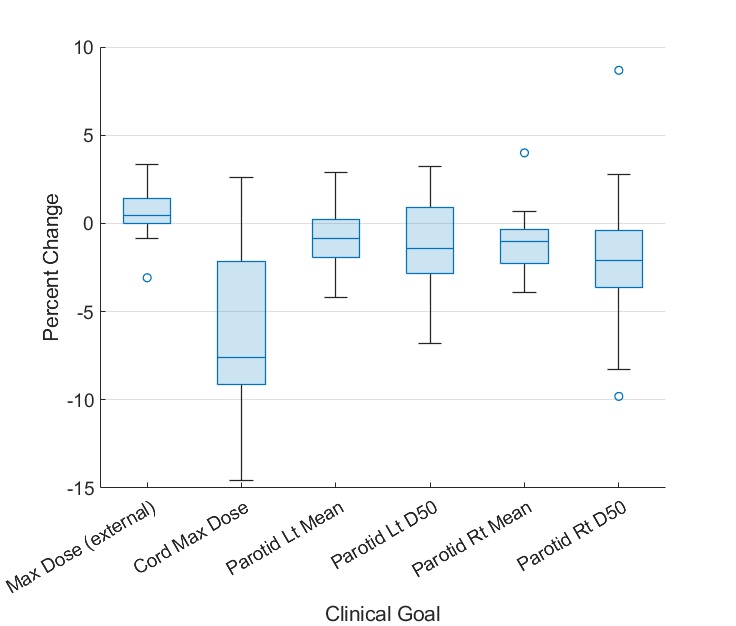}
        \label{fig:predM_c}
    }
    
    \caption{Percent difference of the predicted and mimicked dose distributions}
    \label{fig:PredvMimic}
\end{figure}

For the Tomotherapy mimic plans, only the spinal cord maximum dose was clinically different (>5\%), though in this case the dose was lower in the deliverable plan than the predicted plan. Seven of the nine clinical goals (max dose to external and cord, mean dose and D50 to right and left parotids, and mean larynx dose) were statistically significantly different (p<0.03) than the predicted dose. However, except the external max dose, all the organs-at-risk (OAR) doses were lower in the deliverable Tomotherapy plans than the predicted dose.Figure \ref{fig:predM_c} shows the distribution of the percent difference between the predicted doses and mimicked Tomotherapy plans for select dosimetric values. A full comparison can be found in Supplementary Table \ref{tab:supp_table_5}. Figure \ref{fig:Mimic} shows the dose volume histogram, a common plot for assessing radiation therapy plans that shows the volume of different organs or targets that receive a given dose or above, for a representative patient, demonstrating the similarity in the dose distribution between the predicted and mimicked dose distributions.
\end{multicols}

\begin{figure}[H]
    \centering
    \subfigure[Comparison to the IMRT mimic plan]{
        \includegraphics[width=0.45\textwidth]{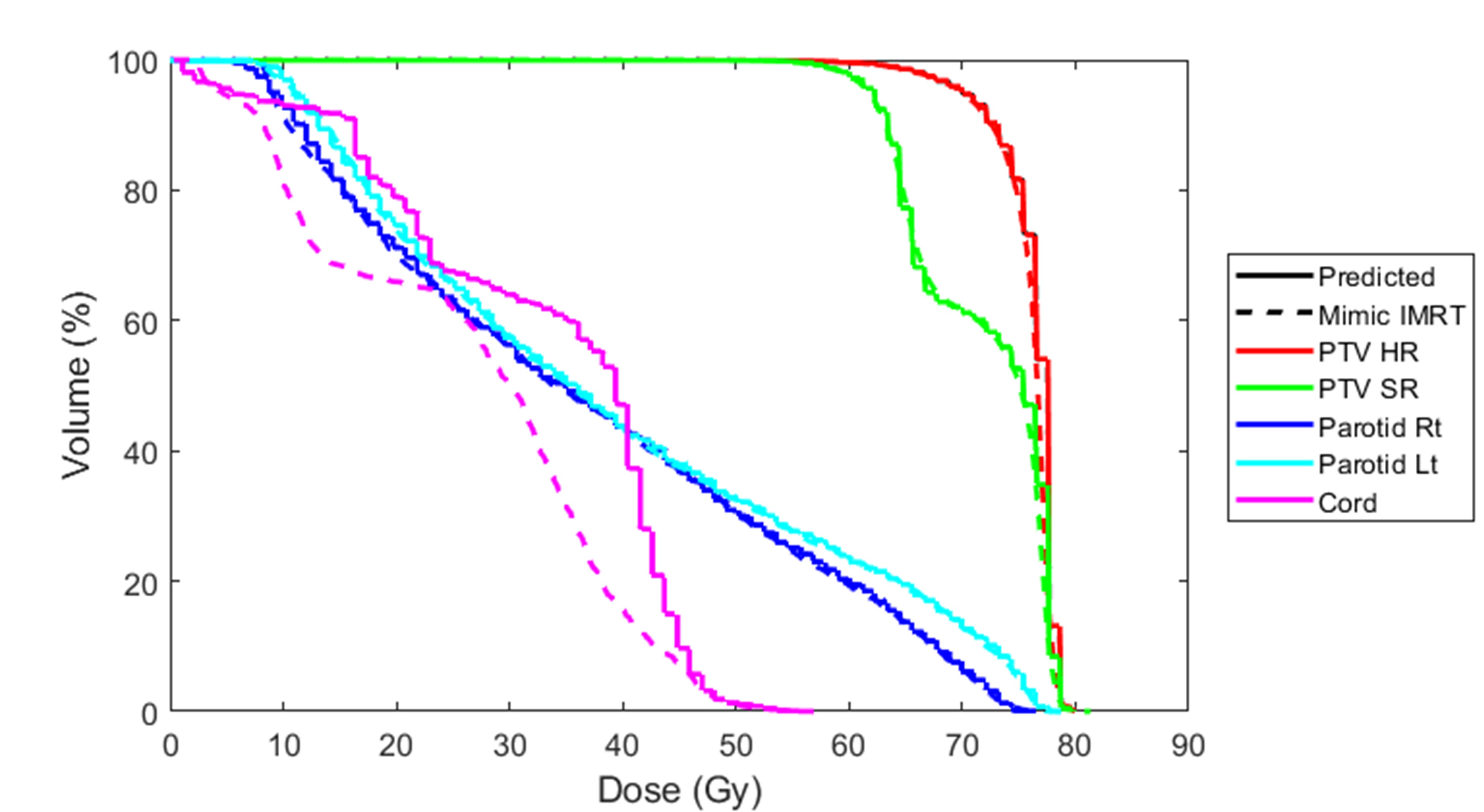}
        \label{fig:mimic_a}
    }
    \subfigure[Comparison to the VMAT mimic plan]{
        \includegraphics[width=0.45\textwidth]{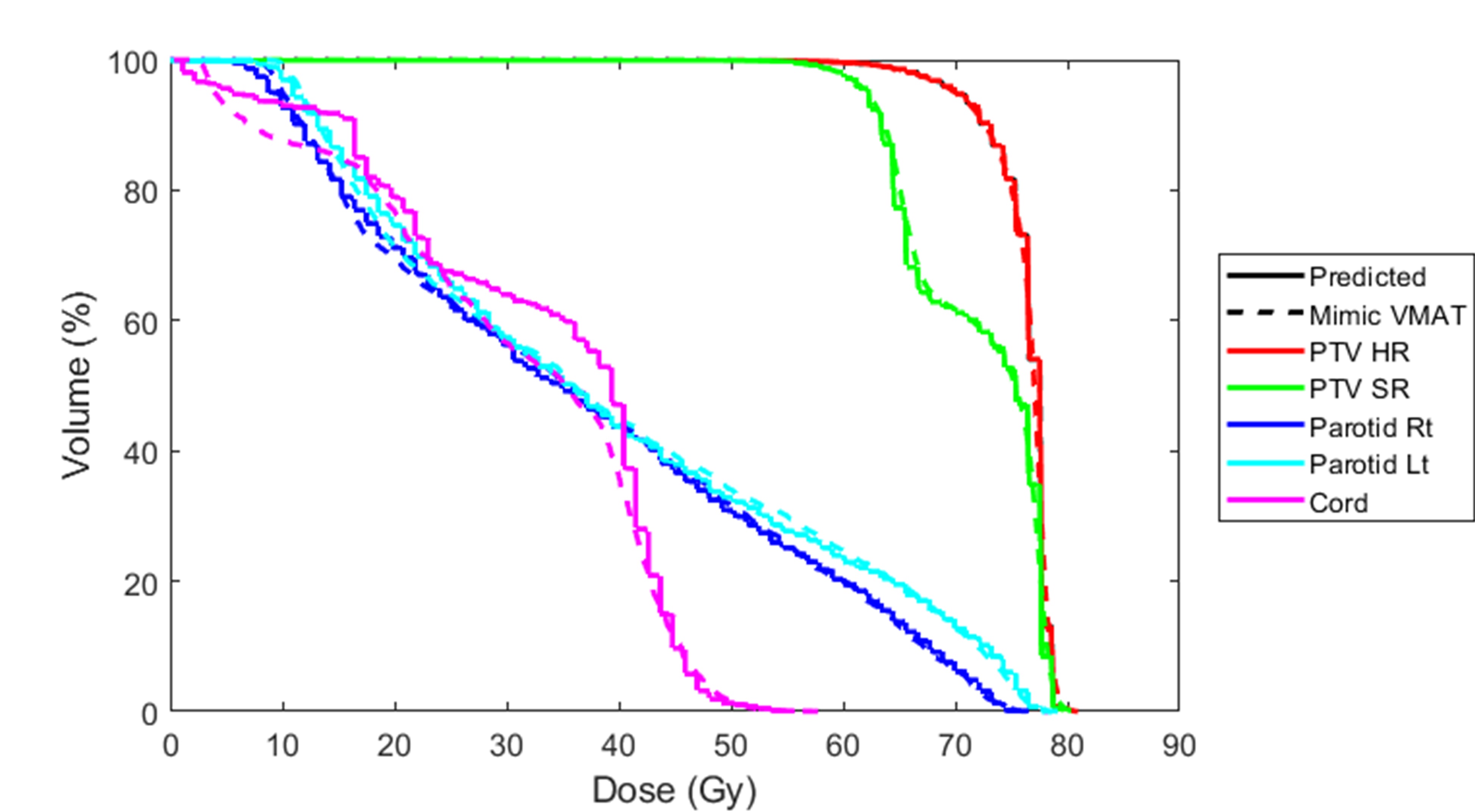}
        \label{fig:mimic_b}
    }
    \subfigure[Comparison to the Tomotherapy mimic plan]{
        \includegraphics[width=0.45\textwidth]{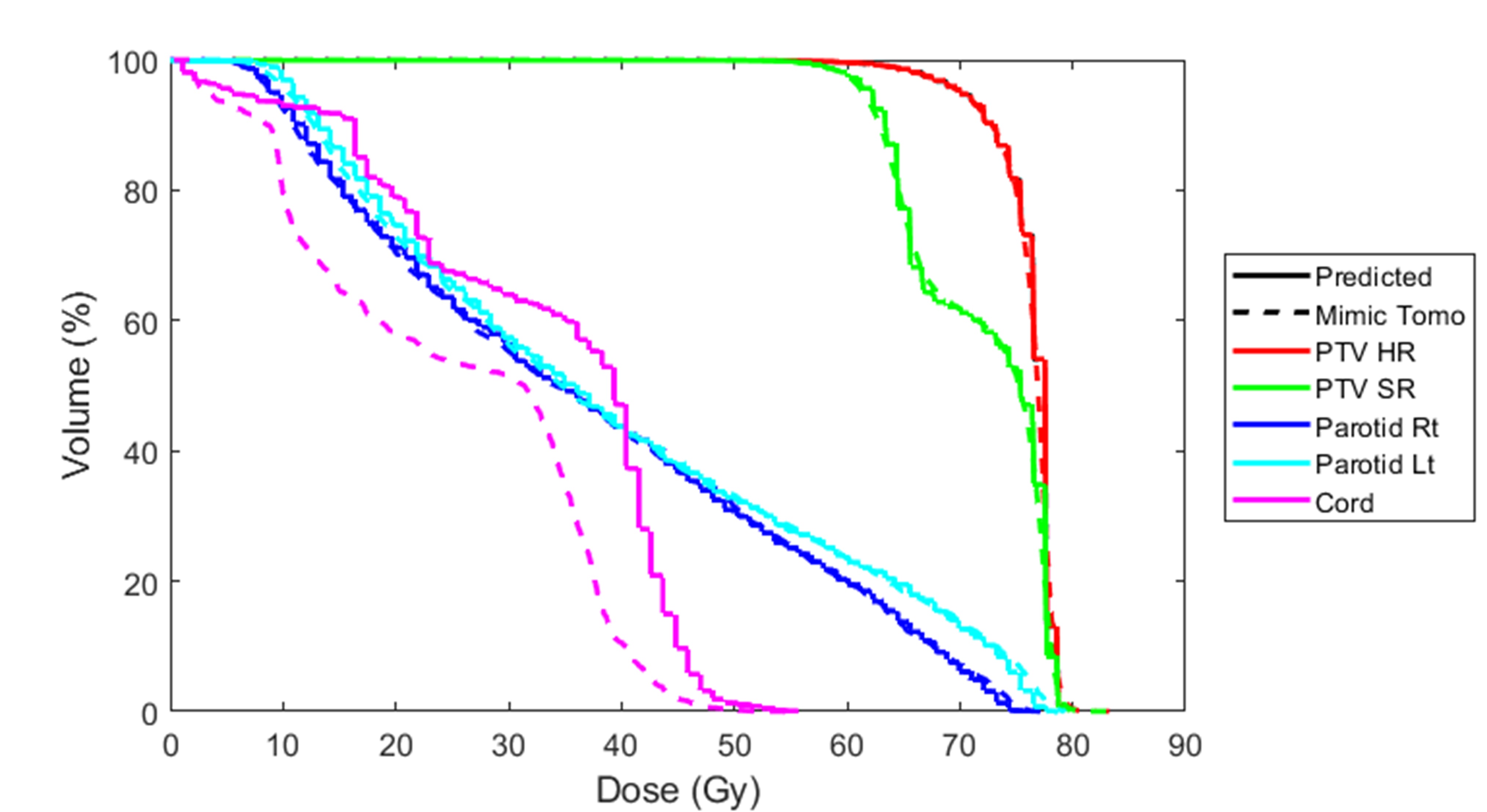}
        \label{fig:mimic_c}
    }
    \subfigure[Comparison of all the mimic plans. All mimic plans achieved good agreement of the target and organ at risk dose volume histograms. ]{
        \includegraphics[width=0.45\textwidth]{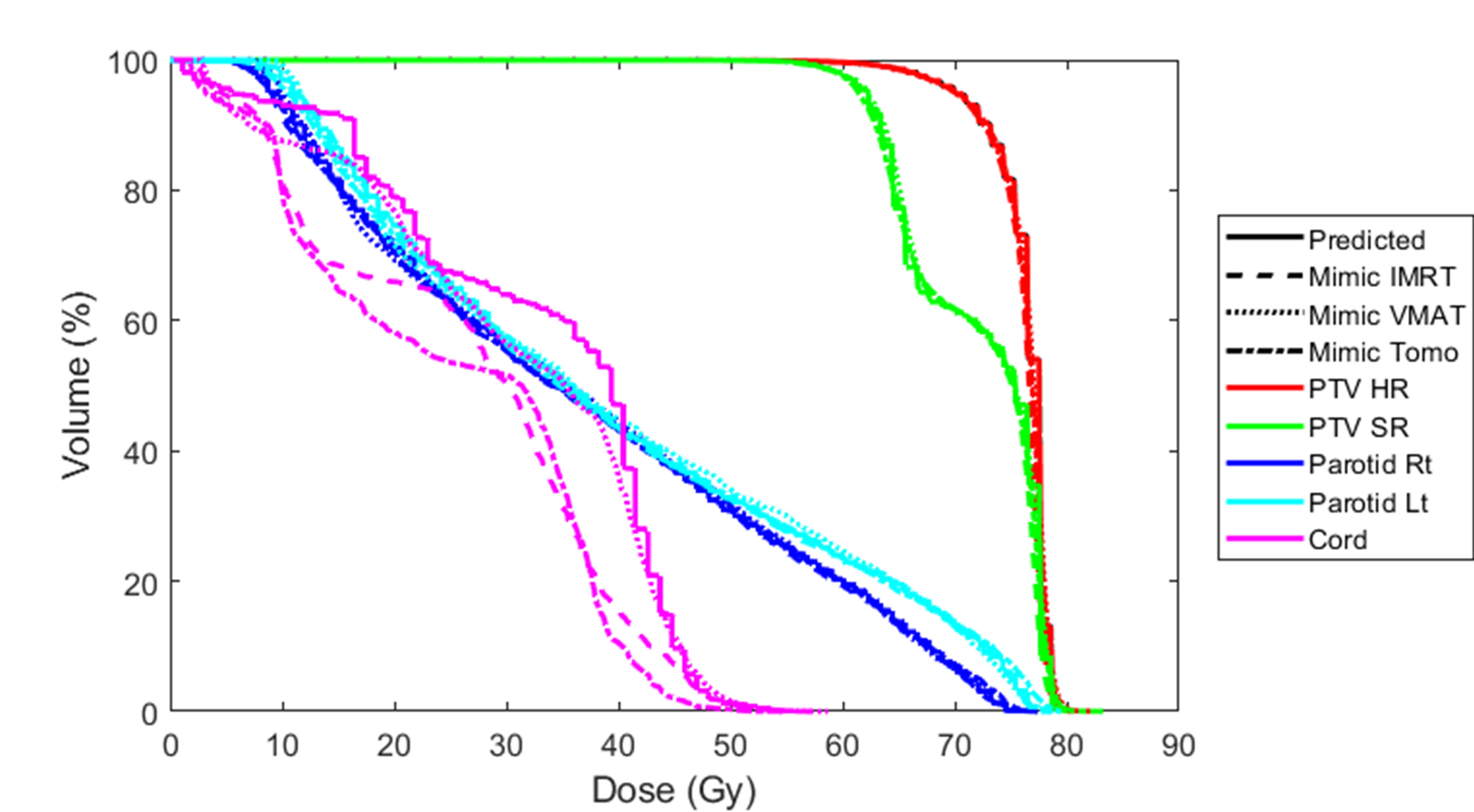}
        \label{fig:mimic_d}
    }
    \caption{Comparison of the dose volume histograms for a representative case. The reference is the predicted dose.}
    \label{fig:Mimic}
\end{figure}

\begin{multicols}{2}

\subsection{Across Protocol}
The integration of the MCO into the clinical plan creation process was found to greatly decrease the dose to OARs (organs at risk) relevant to the clinical goals used at our institution. While the average PTV (Planning Target Volume) standard risk (PTV\_SR) coverage was lower, on average, in the MCO integrated plans, it remained above 95\% for all the cases investigated, as this is the clinical coverage standard used to generate the optimization functions. The difference in clinical goals was found to be clinically relevant (>5\%) and statistically (p<0.03) significant in all cases except for the max dose to the external, and the larynx mean dose. Figure 5 shows the distribution of the percent change of the dosimetric values obtained for select clinical goals between MCO plans and the predicted dose distributions. Figure 6 shows comparisons of the IMRT, VMAT, and Tomotherapy MCO plan DVHs to the prediction. Among the three MCO plans, there was no clinically relevant difference between the IMRT MCO and VMAT MCO plans as assessed by the clinical goals. However, the cord (max) and parotid (mean and D50) doses were significantly lower in the Tomotherapy MCO plans compared to the VMAT MCO plans (p < 0.05). The dose to these structures was on average lower in the Tomotherapy MCO plans compared to the IMRT MCO plans, but the difference did not meet the level of significance (p > 0.05). 
\end{multicols}

\begin{figure}[H]
    \centering
    \includegraphics[width=1\linewidth]{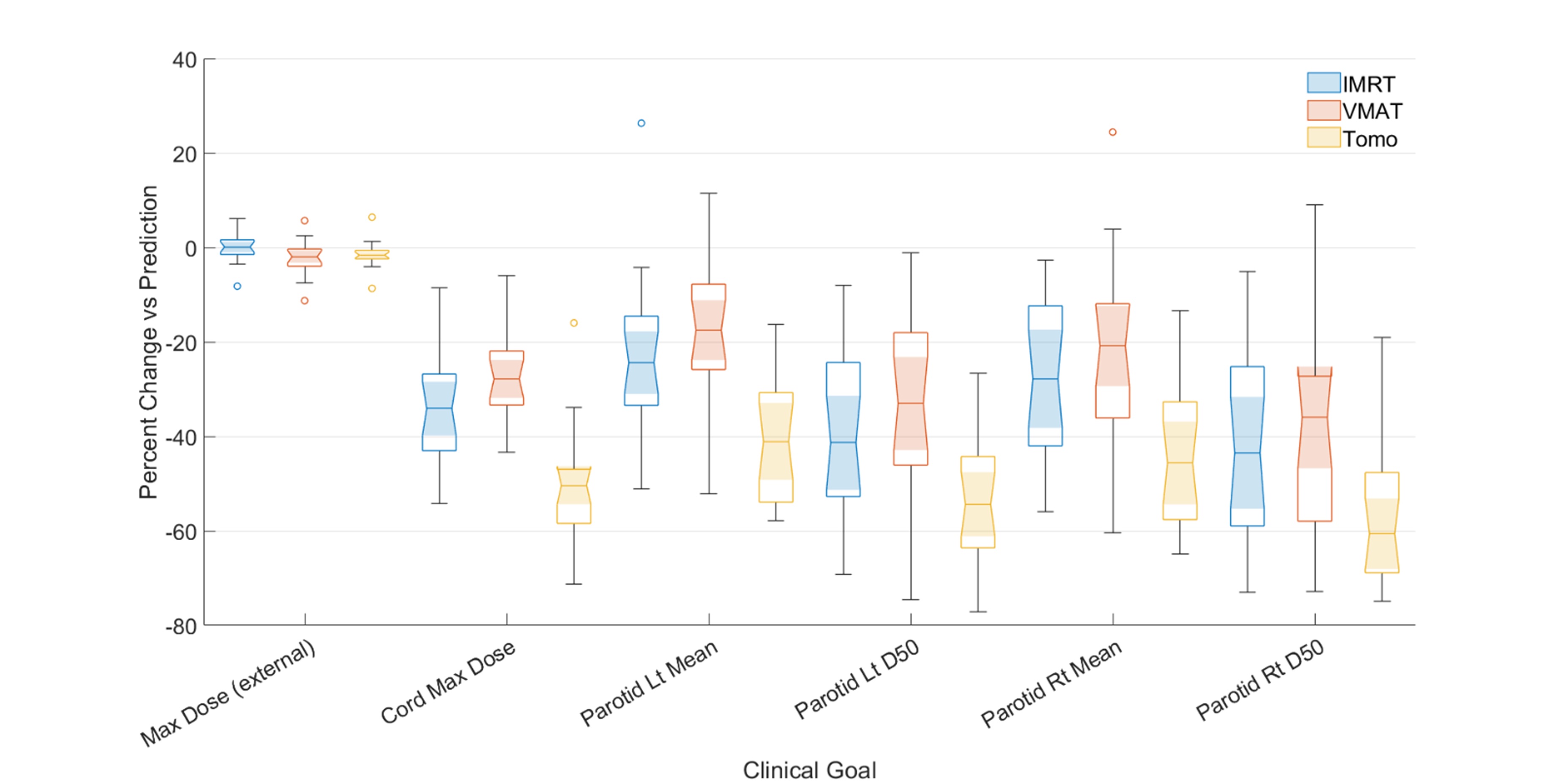}
    \caption{Comparison of MCO and predicted dose distributions
Percent change of the dosimetric values obtained for select clinical goals between the predicted dose distributions and MCO plans. The colors correspond to the treatment technique used in the MCO plan and the shaded region indicates the 95\% confidence interval around the median.
}
    \label{fig:MCOvPredict}
\end{figure}

\begin{figure}[H]
    \centering
    \subfigure[Comparison of the predicted dose to the IMRT MCO plan]{
        \includegraphics[width=0.45\textwidth]{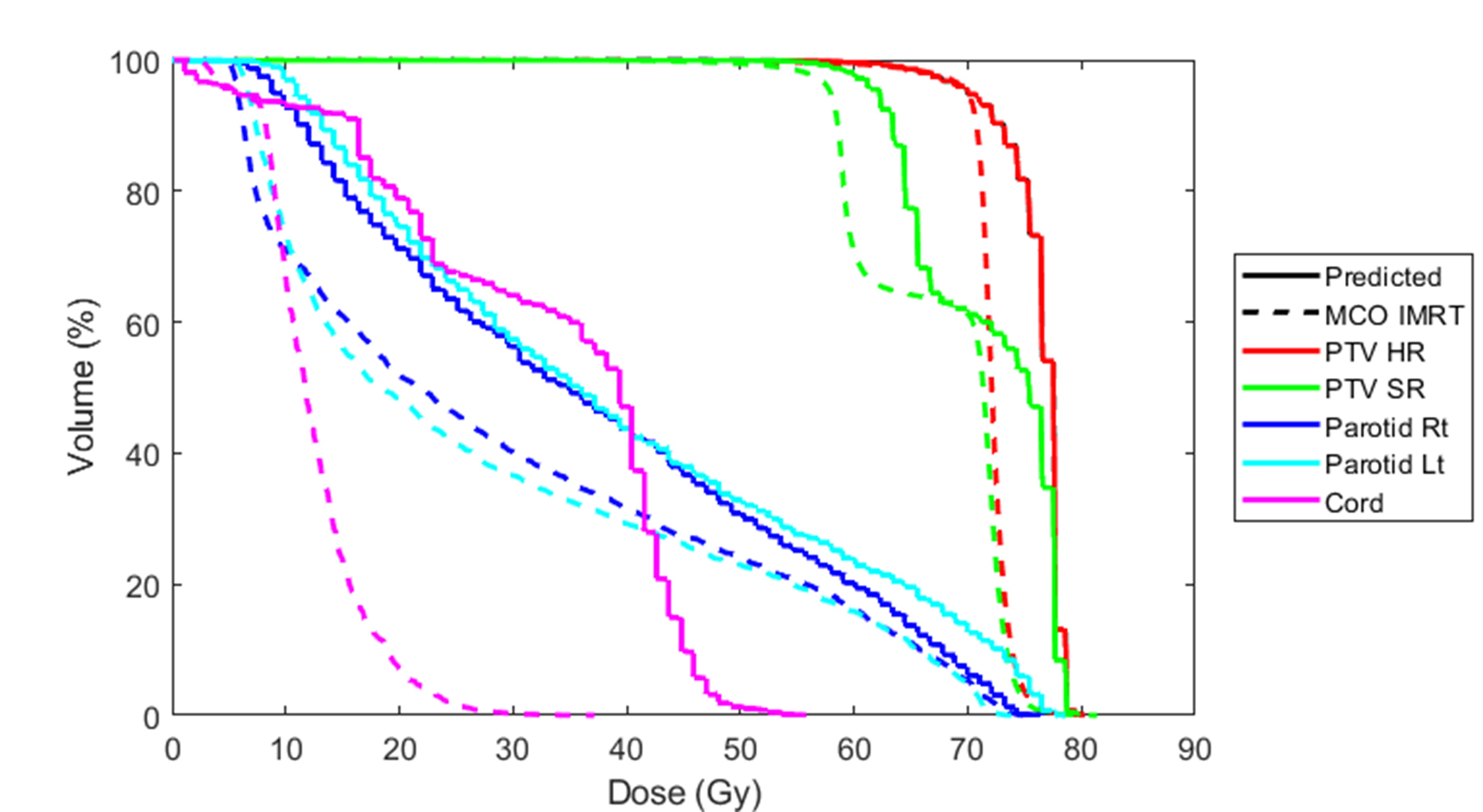}
        \label{fig:MCO_a}
    }
    \subfigure[Comparison to the VMAT MCO plan]{
        \includegraphics[width=0.45\textwidth]{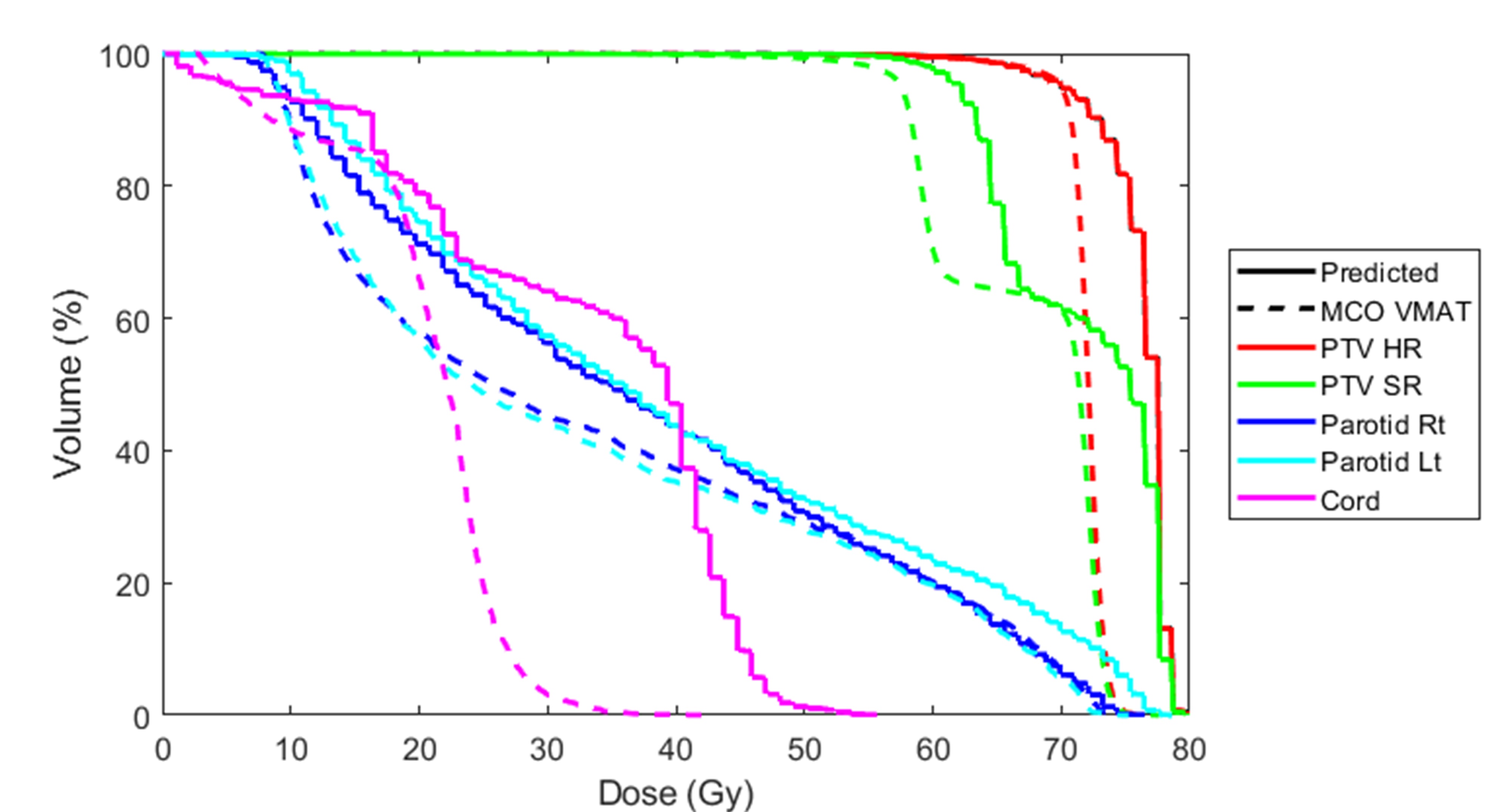}
        \label{fig:MCO_b}
    }
    \subfigure[Comparison to the Tomotherapy MCO plan]{
        \includegraphics[width=0.45\textwidth]{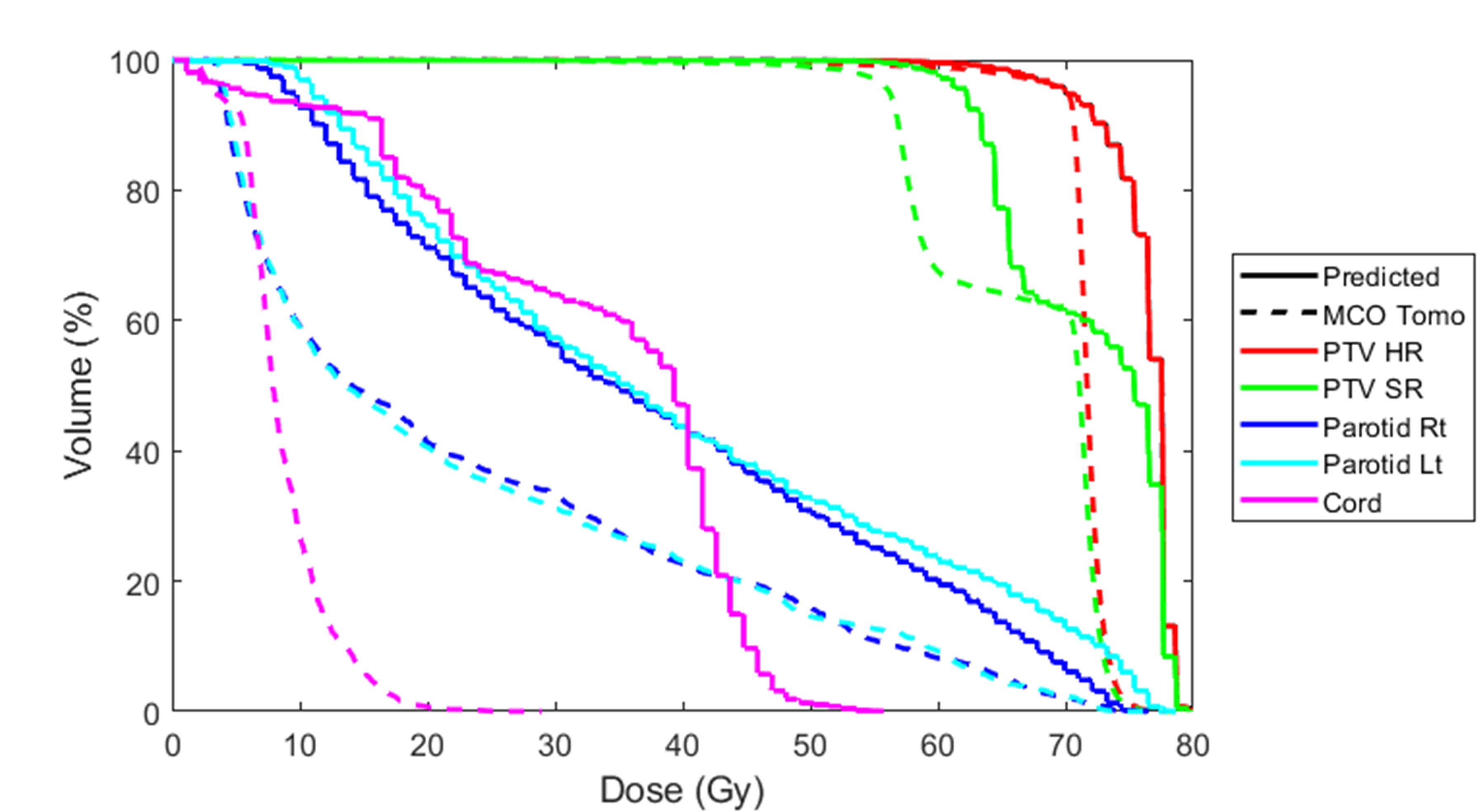}
        \label{fig:MCO_c}
    }
    \subfigure[Comparison of all the MCO plans to the predicted dose. In all cases, the uniformity of the PTV doses at the prescription dose was increased, allowing for a lower total dose and substantial reduction in the dose to the parotids and the spinal cord]{
        \includegraphics[width=0.45\textwidth]{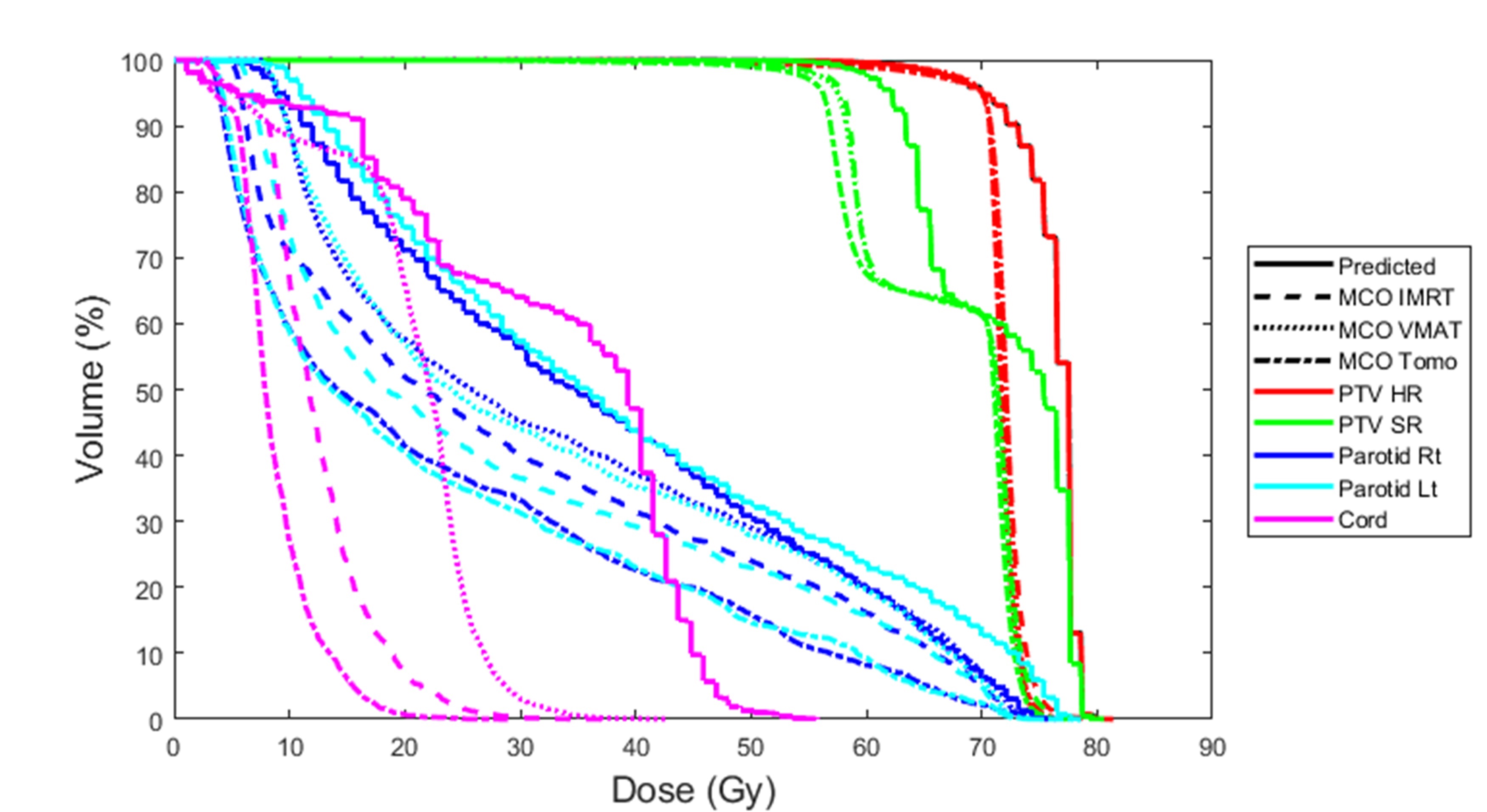}
        \label{fig:MCO_d}
    }
    \caption{Comparison of the dose volume histograms for a representative case, MCO plans}
    \label{fig:MCO}
\end{figure}

\begin{multicols}{2}

\subsection{Delivery QA Testing}
Delivery QA for all cases passed above 98\% (range 98-100\%) using a gamma criteria of 3\%/2mm, well above the 90\% threshold recommendation by the American Association of Physicists in Medicine \cite{Ref39}. As such, the mimicked and MCO plans for all three modalities were considered clinically deliverable.

\section{Discussion}
We successfully developed a workflow using dose mimicking optimization to rapidly create deliverable plans matching the dose distributions predicted by our model, utilizing both the same modality as in the plans used for training the model, IMRT, as well as for different modalities, VMAT and Tomotherapy. Based on the clinical goals, the predicted dose distributions produced by our model are deliverable and clinically acceptable across all modalities investigated. The noted difference in the external and cord maximum dose between the predicted doses and the mimicked plans is likely explained by the optimization process; as the mimic dose operation attempts to match the DVH (dose volume histograms) and the voxel-by-voxel dose of the predicted dose, it is understandable that the max dose, which is defined by both a small portion of the DVH and a small number of the voxels, could be missed in the optimization process. We found that the inclusion of maximum dose optimization objectives for the cord and external based on the prediction reduced the difference of the maximum dose to these structures below clinically relevant levels (<5\%) while having little to no effect on the other clinical goals. As such, we would recommend explicitly adding optimization objectives based on the prediction for important, small volume dose points.  More importantly, since the predicted dose distributions are clinically accurate across a variety of treatment modalities, it is potentially feasible to create a single trained model that could be used for plan generation across multiple treatment modalities.
Integrating the multi-criteria optimizer into the workflow further allowed quick adjustments to reflect institutional preference, opening the possibility of extending a single model across multiple institutions. Additionally, this makes it easier to tailor a dose distribution to edge cases that would not be well captured in the training set, such as patients with prior radiation that have lower than usual dose tolerance to certain OARs. While it is a powerful tool, the user still needs to define proper objectives and constraints to assure the pareto-surface covers the full area of potential interest while restricting it to only clinically acceptable plans to increase the ease of arriving at a high-quality plan \cite{Ref40}. Our model serves as a useful way to assist with these definitions. While in this work only the parotid D50 and the external and spinal cord max doses were set as hard constraints, it would be possible to set hard constraints for all OARs of interest, although the dose constraints used would likely have to be slightly raised to give some room to modify the dose distribution. The MCO could also be used to improve the dosimetry for specific patients compared to the dose prediction, as the prediction is based on the optimal results derived from the training samples. In these cases, the dose prediction would serve as a good starting point for the new plan, and the planner could use the MCO functionality to assess if any further dosimetric improvement or preference adjustment is possible. As with other supervised learning models, these improved plans could be incorporated into the training set to iteratively improve the performance of the model, by further increasing the quality of the training plans and/or including more that match an institution’s specific preference. 
Our work on the MCO approach did show that a prediction model is less accurate when the current plan and training plans have differing dosimeric preferences, as seen by the significant difference in most of the clinical goals seen in section 3.4. There was also a slight decrease in the average coverage of the targets at prescription dose, from 97\% to 95\%. It is likely that this is due to the use of 95\% coverage in the optimization goals, based on our clinical preference.  Nevertheless, employing the MCO approach may contribute to additional sparing of OARs or adapting the prediction to align with alternative clinical preferences. In our study, we observed average reductions nearing 45\% for certain structures when compared to the initial test plan based on predictions, specifically for dose metrics that hold significance at our institution (refer to Supplementary Table 6). This is a common problem with supervised learning models, as the trained model will not deviate far from the data on which it was trained. While the predicted doses did seem to help give a reasonable estimate of the maximum cord and external dose that could be achieved, it is likely the integration with MCO is helpful to optimize the dosimetry closer to the desired protocol. As mentioned above, the plans generated with this approach could be included in the training set to modify the model to better match institutional preference, should a center have the resources to do so. However, for lower-resource centers, the prediction can serve as a good starting point and reference during the MCO process.  Additionally, our dose prediction model was unable to predict dose distributions directly for plans with multiple targets where the target prescription ratios are different from those used in the training set. While changes in the overall scaling can be handled by simply renormalizing the isodose distribution to the desired prescription, if the ratio between target doses changes then the isodose distribution must be adjusted, most significantly in the gradient between targets. 

Recently developed reinforcement learning approaches \cite{Ref41, Ref42, Ref43} offer one alternative that can allow for a way to reduce the overhead involved in transferring a model between institutions. The reward function used to assess the plans can be modified to suit the specific preference of an institution and relatively few patient cases (10s instead of 100s) are needed for training. However, a new model does need to be trained each time the reward function is modified, which would slow deployment at a new center. Further, the current architectures do not scale well with the number of relevant organs at risk (of which there are many for head and neck cases), though work is being done to minimize this issue \cite{Ref41}.
Future work will focus on methods to allow for the model to be applicable across multiple prescription levels, which would extend the flexibility of the model. Further work could also be done to develop an efficient method to modify a standardized trained model to conform to the specific preference of an institution. This can help deploy individualized models to medical centers without sufficient resources to build a model on their own.

\section{Conclusion}
Our deep-learning-based model can create predicted dose distributions that are deliverable with a variety of radiotherapy treatment modalities, including those not used to create the plans used for training the model. Integrating the multi-criteria optimizer allows for customization of the predicted plan with only a minor increase in the planning time.
\end{multicols}
\clearpage

\newpage

\bibliographystyle{ieeetr}
\bibliography{references}

\newpage

\appendix
\section*{Supplementary Material}
\captionsetup[table]{name=Supplementary Table}

\begin{table}[h]
\centering
\caption{Comparison of predicted vs mimicked dose distributions using IMRT}
\begin{tabular}{lccc}
\toprule
Clinical Goal & Mimicked-Predicted & \% difference & p-value \\
\midrule
PTV\_SR Coverage & -0.12 $\pm$ 0.37 \% & -0.12 $\pm$ 0.39 & 0.43 \\
External Max Dose & 0.8 $\pm$ 1.0 Gy & 1.11 $\pm$ 1.25 & 0.00* \\
Cord Max Dose & 2.1 $\pm$ 2.7 Gy & 5.57 $\pm$ 6.97 & 0.00* \\
Lt Parotid Mean & 0.0 $\pm$ 0.9 Gy & 0.42 $\pm$ 4.08 & 0.77 \\
Lt Parotid D50 & 0.1 $\pm$ 1.5 Gy & 1.58 $\pm$ 6.03 & 0.62 \\
Rt Parotid Mean & 0.1 $\pm$ 0.7 Gy & 0.45 $\pm$ 1.93 & 0.22 \\
Rt Parotid D50 & -0.6 $\pm$ 1.0 Gy & -1.52 $\pm$ 3.11 & 0.05 \\
Larynx Mean & -0.4 $\pm$ 1.2 Gy & -1.69 $\pm$ 3.84 & 0.30 \\
Larynx V60 & 1.13 $\pm$ 1.08 \% & 5.88 $\pm$ 7.97 & 0.38 \\
\bottomrule
\multicolumn{4}{l}{*statistically significant, p<0.05} \\
\end{tabular}
\label{tab:sup_tab_1}
\end{table}

\begin{table}[h]
\centering
\caption{Comparison of predicted vs mimicked dose distributions using VMAT}
\begin{tabular}{lccc}
\toprule
Clinical Goal & Mimicked-Predicted & \% difference & p-value \\
\midrule
PTV\_SR Coverage & 0.04 $\pm$ 0.51 \% & 0.04 $\pm$ 0.53 & 0.21 \\
External Max Dose & 0.7 $\pm$ 1.3 Gy & 0.89 $\pm$ 1.67 & 0.00* \\
Cord Max Dose & 1.0 $\pm$ 1.7 Gy & 2.83 $\pm$ 4.84 & 0.04* \\
Lt Parotid Mean & 0.0 $\pm$ 0.8 Gy & 0.68 $\pm$ 2.92 & 0.49 \\
Lt Parotid D50 & 0.0 $\pm$ 1.1 Gy & 1.49 $\pm$ 5.95 & 0.99 \\
Rt Parotid Mean & 0.1 $\pm$ 0.8 Gy & 0.55 $\pm$ 2.08 & 0.38 \\
Rt Parotid D50 & -0.2 $\pm$ 1.0 Gy & -0.16 $\pm$ 2.39 & 0.83 \\
Larynx Mean & -0.3 $\pm$ 1.4 Gy & -1.53 $\pm$ 4.64 & 0.91 \\
Larynx V60 & -0.27 $\pm$ 0.85 \% & -2.23  $\pm$ 9.98 & 0.33 \\
\bottomrule
\multicolumn{4}{l}{*statistically significant, p<0.05} \\
\end{tabular}
\label{tab:sup_tab_2}
\end{table}

\begin{table}[h]
\centering
\caption{Comparison of predicted vs mimicked dose distributions using Tomotherapy}
\begin{tabular}{lccc}
\toprule
Clinical Goal & Mimicked-Predicted & \% difference & p-value \\
\midrule
PTV\_SR Coverage & 0.06 $\pm$ 0.33 \% & 0.06 $\pm$ 0.34 & 0.33 \\
External Max Dose & 0.6$\pm$ 1.4 Gy & 0.72 $\pm$ 1.84 & 0.03* \\
Cord Max Dose & -2.6 $\pm$ 1.8 Gy & -6.29 $\pm$ 4.12 & 0.00* \\
Lt Parotid Mean & -0.5 $\pm$ 0.8 Gy & -1.24 $\pm$ 1.46 & 0.01* \\
Lt Parotid D50 & -0.6 $\pm$ 0.9 Gy & -1.74 $\pm$ 2.86 & 0.02* \\
Rt Parotid Mean & -0.7 $\pm$ 0.7 Gy & -1.62 $\pm$ 1.53 & 0.01* \\
Rt Parotid D50 & -1.1 $\pm$ 1.2 Gy & -2.44 $\pm$ 2.24 & 0.03* \\
Larynx Mean & -1.4 $\pm$ 1.0 Gy & -2.89 $\pm$ 2.35 & 0.02* \\
Larynx V60 & -1.23 $\pm$ 1.93\% & -9.35 $\pm$ 24.46 & 0.38 \\
\bottomrule
\multicolumn{4}{l}{*statistically significant, p<0.05} \\
\end{tabular}
\label{tab:sub_tab_3}
\end{table}

\begin{table}[h]
\centering
\caption{Summary comparison of predicted vs mimicked dose distributions for various modalities}
\begin{tabular}{lcccccc}
\toprule
 & \multicolumn{2}{c}{IMRT} & \multicolumn{2}{c}{VMAT} & \multicolumn{2}{c}{Tomotherapy} \\
\cmidrule(lr){2-3} \cmidrule(lr){4-5} \cmidrule(lr){6-7}
Clinical Goal & \% difference & p-value & \% difference & p-value & \% difference & p-value \\
\midrule
PTV\_SR Coverage & 0.24 $\pm$ 0.21 & 0.43 & 0.26 $\pm$ 0.20 & 0.21 & 0.23 $\pm$ 0.28 & 0.33 \\
External Max Dose & 1.25 $\pm$ 0.88 & 0.00* & 1.48 $\pm$ 1.05 & 0.00* & 1.35 $\pm$ 1.14 & 0.03* \\
Cord Max Dose & 6.58 $\pm$ 4.49 & 0.00* & 3.46 $\pm$ 2.81 & 0.04* & 6.02 $\pm$ 4.28 & 0.00* \\
Lt Parotid Mean & 2.47 $\pm$ 2.91 & 0.77 & 2.46 $\pm$ 2.21 & 0.49 & 1.48 $\pm$ 1.32 & 0.01* \\
Lt Parotid D50 & 3.82 $\pm$ 4.31 & 0.62 & 4.15 $\pm$ 4.79 & 0.99 & 2.27 $\pm$ 2.23 & 0.02* \\
Rt Parotid Mean & 1.51 $\pm$ 1.52 & 0.22 & 1.95 $\pm$ 1.16 & 0.38 & 1.88 $\pm$ 1.37 & 0.01* \\
Rt Parotid D50 & 2.37 $\pm$ 2.64 & 0.05 & 1.96 $\pm$ 0.97 & 0.83 & 2.46 $\pm$ 2.37 & 0.03* \\
Larynx Mean & 0.98 $\pm$ 0.52 & 0.30 & 0.53 $\pm$ 0.38 & 0.91 & 2.89 $\pm$ 2.35 & 0.02* \\
Larynx V60 & 6.66 $\pm$ 7.1 & 0.38 & 7.18  $\pm$ 8.50 & 0.33 & 15.04  $\pm$ 20.33 & 0.38 \\
Other Tissue Mean & 8.35 $\pm$ 5.92 & 0.00* & 10.54 $\pm$ 4.35 & 0.00* & 7.18 $\pm$ 3.63 & 0.00* \\
\bottomrule
\multicolumn{4}{l}{*statistically significant, p<0.05} \\
\end{tabular}
\label{tab:sup_table_4}
\end{table}

\begin{table}[H]
\caption{Comparison of predicted dose vs plans created using MCO method}
\centering
\begin{tabular}{lccc}
\toprule
\textbf{Clinical Goal} & \multicolumn{3}{c}{\textbf{Difference (\%)}} \\
\cmidrule{2-4}
 & \textbf{IMRT} & \textbf{VMAT} & \textbf{Tomotherapy} \\
\midrule
PTV\_SR Coverage & $-1.1 \pm 1.0\%$ & $-0.8 \pm 1.1\%$ & $-2.5 \pm 1.0\%$ \\
External Max Dose & $0.0 \pm 2.5$ Gy & $-1.7 \pm 3.0$ Gy & $-1.2 \pm 2.2$ Gy \\
Cord Max Dose & $-14.0 \pm 6.3$ Gy & $-11.9 \pm 5.9$ Gy & $-21.4 \pm 6.7$ Gy \\
Lt Parotid Mean & $-8.1 \pm 4.6$ Gy & $-6.6 \pm 5.1$ Gy & $-14.0 \pm 4.8$ Gy \\
Lt Parotid D50 & $-13 \pm 6.2$ Gy & $-12.1 \pm 7.9$ Gy & $-18.0 \pm 7.2$ Gy \\
Rt Parotid Mean & $-9.3 \pm 5.1$ Gy & $-7.7 \pm 6.0$ Gy & $-15 \pm 4.8$ Gy \\
Rt Parotid D50 & $-13.6 \pm 5.6$ Gy & $-12.5 \pm 6.8$ Gy & $-18.6 \pm 6.1$ Gy \\
Larynx Mean & $-3.3 \pm 8.8$ Gy & $-6.0 \pm 8.2$ Gy & $-7.6 \pm 10.7$ Gy \\
Larynx V60 & $1.7 \pm 14.7\%$ & $-6.2 \pm 11.0\%$ & $-3.0 \pm 14.5\%$ \\
Other Tissue Mean & $3.1 \pm 1.7$ Gy & $2.9 \pm 1.4$ Gy & $3.1 \pm 2.1$ Gy \\
\bottomrule
\end{tabular}
\label{tab:supp_table_5}
\end{table}

\begin{table}[h]
\caption{Tradeoff constraints and objectives used for MCO integration}
\centering
\begin{tabular}{ll}
\toprule
\textbf{ROI} & \textbf{Goal} \\
\midrule
\multicolumn{2}{l}{\textit{Constraints}} \\
High Risk PTV & Volume receiving 70 Gy $\geq 95\%$ \\
Intermediate Risk PTV* & Volume receiving 63 Gy $\geq 95\%$ \\
Standard Risk PTV & Volume receiving 56 Gy $\geq 95\%$ \\
Spinal Cord & Max: X Gy point dose \\
External & Max: X Gy point dose \\
Parotid Rt/Lt & 50\% receives $<$ X Gy \\
\midrule
\multicolumn{2}{l}{\textit{Objectives}} \\
High Risk PTV & Uniform dose 70 Gy \\
 & Volume receiving 70 Gy = 100\% \\
Intermediate Risk PTV* & Volume receiving 63 Gy = 100\% \\
Standard Risk PTV & Volume receiving 56 Gy = 100\% \\
Intermediate Risk PTV – High Risk PTV* & Max: 63 Gy point dose \\
Standard Risk PTV – High Risk PTV & Max: 56 Gy point dose \\
External & Dose fall-off: 56 Gy to 46 Gy in 0.5 cm \\
 & Dose fall-off 56 Gy to 5 Gy in 5 cm \\
 & Max: 70.5 Gy point dose \\
Parotid Rt/Lt & Max EUD: 1 Gy, A = 1 \\
Spinal Cord & Max: 5 Gy point dose \\
Larynx* & Max EUD: 10 Gy A = 1 \\
 & Max: 50 Gy point dose \\
Esophagus* & Volume receiving 1 Gy $<$ 33\% \\
 & Volume receiving 1 Gy $<$ 66\% \\
\bottomrule
\multicolumn{2}{l}{*When applicable} \\
\end{tabular}
\label{tab:supp_table_6}
\end{table}

\begin{table}[h]
\caption{Target and OAR constraints used at our institution for head and neck planning}
\centering
\begin{tabular}{ll}
\toprule
\textbf{ROI} & \textbf{Clinical Goal} \\
\midrule
High Risk PTV & Volume receiving 70 Gy $\geq 95\%$ \\
Intermediate Risk PTV* & Volume receiving 63 Gy $\geq 95\%$ \\
Standard Risk PTV* & Volume receiving 56 Gy $\geq 95\%$ \\
Spinal Cord & Max: 50 Gy to point dose \\
Parotid Rt/Lt & Mean dose $<$ 26 Gy \\
 & 50\% receives $<$ 30 Gy \\
Larynx & Mean dose $\leq$ 41 Gy \\
 & Volume receiving 60 Gy $\leq$ 24\% \\
External (all tissue) & Max: 77 Gy to point dose \\
\bottomrule
\multicolumn{2}{l}{*When applicable} \\
\end{tabular}
\label{tab:supp_table_7}
\end{table}

\end{document}